\documentclass[pra,twocolumn,footinbib]{revtex4}
\usepackage{amssymb,amsfonts,amsmath,dcolumn,bm}
\usepackage[dvips]{graphicx}

\newcommand \beq{\begin{eqnarray}}
\newcommand \eeq{\end{eqnarray}}

\newcommand{\feynslash}[1]{{#1\kern-.5em /}}
\def\simge{\mathrel{%
         \rlap{\raise 0.511ex \hbox{$>$}}{\lower 0.511ex \hbox{$\sim$}}}}
\def\simle{\mathrel{
         \rlap{\raise 0.511ex \hbox{$<$}}{\lower 0.511ex \hbox{$\sim$}}}}

\begin{document}

\title{Two-slit diffraction with highly charged particles:  Niels Bohr's
consistency argument that the electromagnetic field must be quantized}
\author{Gordon Baym}
\email{gbaym@illinois.edu}
\affiliation{Department of Physics, University of Illinois at Urbana-Champaign, 1110 W. Green Street, Urbana, Illinois 61801}
\author{Tomoki Ozawa}
\email{tozawa2@illinois.edu}
\affiliation{Department of Physics, University of Illinois at Urbana-Champaign, 1110 W. Green Street, Urbana, Illinois 61801}

\begin{abstract}

    We analyze Niels Bohr's proposed two-slit interference experiment with
highly charged particles that argues that the consistency of elementary
quantum mechanics requires that the electromagnetic field must be quantized.
In the experiment a particle's path through the slits is determined by
measuring the Coulomb field that it produces at large distances; under these
conditions the interference pattern must be suppressed.  The key is that as
the particle's trajectory is bent in diffraction by the slits it must radiate
and the radiation must carry away phase information.  Thus the radiation field
must be a quantized dynamical degree of freedom.  On the other hand, if one
similarly tries to determine the path of a massive particle through an
inferometer by measuring the Newtonian gravitational potential the particle
produces, the interference pattern would have to be finer than the Planck
length and thus undiscernable.  Unlike for the electromagnetic field, Bohr's
argument does not imply that the gravitational field must be quantized.

\end{abstract}

\maketitle

    Niels Bohr once suggested a very simple \textit{gedanken} experiment to
prove that, in order to preserve the consistency of elementary quantum
mechanics, the radiation field must be quantized as photons \cite{footnote1}.
 In the experiment one carries out conventional two-slit diffraction
with electrons (or other charged particles), building up the diffraction
pattern one electron at a time (as in the experiment of Ref.~\cite{tonomura}).
One then tries to determine which slit the electron went through by measuring
far away, in the plane of the slits, the Coulomb field of the electron as it
passes through the slits.  See Fig.~1.  If the electron passes through the
upper slit it produces a stronger field than if it passes through lower slit.
Thus if one can measure the field sufficiently accurately one gains
``which-path" information, posing the possibility of seeing interference while
at the same time knowing the path the electron takes, a fundamental violation
of the principles of quantum mechanics \cite{footnote2}.

\begin{figure}[b]
\begin{center}
\includegraphics[width=8cm,keepaspectratio=true]{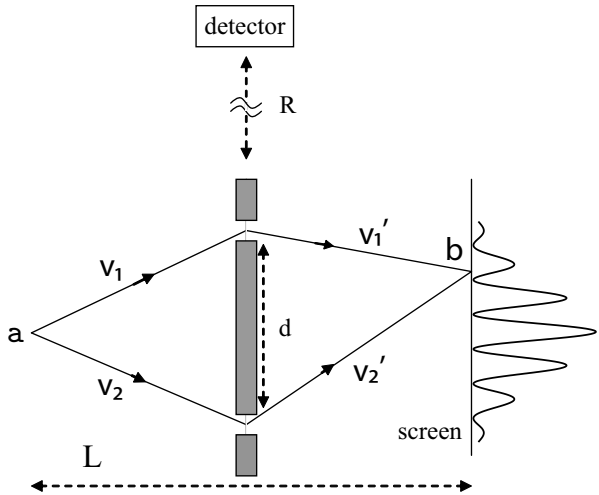}
\end{center}
\caption{Two slit diffraction with single electrons, in which one measures the
Coulomb field produced by the electrons at the far-away detector.
}
\label{setup}
\end{figure}

    In an experiment with ordinary electrons of charge $e$ the uncertainty
principle prevents measurement of the Coulomb field to the required accuracy,
as we shall see below, following the prescription of Bohr and Rosenfeld for
measuring electromagnetic fields \cite{BR1933, BR1950}.  However, as Bohr
pointed out, one can imagine carrying out the same experiment with ({\it
super}) electrons of arbitrarily large charge, $Ze$, and indeed, for
sufficiently large $Z$, one can determine which slit each electron went
through.  However, elementary quantum mechanics requires that once one has the
capability of obtaining which-path information, even in principle, the
interference pattern must be suppressed, independent of whether one actually
performs the measurement.

    Underlying the loss of the pattern is that the electron not only carries a
Coulomb field, but also produces a radiation field as it "turns the corner"
when passing through the slits.  The larger the charge the stronger is the
radiation
produced.  This radiation must introduce a phase uncertainty in order to
destroy the pattern, and so itself must carry phase information; thus the
electromagnetic field must have independent quantum degrees of freedom.  Were
the quantum mechanical electrons to emit classical radiation, the emission
would produce a well-defined phase shift of the electron amplitudes along the
path, which while possibly shifting the pattern, as in the Aharonov-Bohm
effect \cite{aharonov-bohm}, would not destroy it.  In a sense the suppression
of the pattern is an extension of the Aharonov-Bohm effect to fluctuating
electromagnetic potentials (discussed by Aharonov and
Popescu \cite{footnote3}).

    Our object in this paper is to carry out a detailed analysis of the
physics implicit in Bohr's suggested experiment.  After describing the
experiment more fully, we determine the strength of charge needed to measure
the Coulomb field at large distances sufficiently accurately.  We then analyze
how coupling of the particle to the quantized electromagnetic field in
diffraction suppresses the interference pattern, with increasing charge,
before Coulomb measurements can yield which-path information.

    The first experiment that revealed effects of quantization of the
electromagnetic field in interference is that of Grangier et al.
\cite{aspect}, which showed how interference of single photons differs from
classical interference.  The loss of particle coherence in interferometry due
to photon emission was first demonstrated by Pfau et al.  \cite{pfau}, and due
to photon scattering by Chapman et al \cite{chapman}.  Various works, both
theoretical and experimental, have discussed determining the path of charged
particles in the double-slit problem, but none, it seems, in connection with
Bohr's proposed experiment.  The theoretical possibility of distinguishing
paths by measurement of the photon field is discussed in Ref.~\cite{scully},
while Refs.~\cite{furry} and \cite{popescurmp} discuss determining the path
through detection of the electric field inside the loop of the paths.  See
also Stern et al.~\cite{stern} on decoherence due to the interaction of
charged particles with the gauge field.  Experimental attempts to measure
which-path information using interferometers fabricated in high-mobility
two-dimensional electron gases include Refs.~\cite{schuster,buks,chang}.

    A natural question to ask is whether by measuring the Newtonian
gravitational field produced by the mass of a particle as it diffracts, one
can similarly gain which-path information; as we show, the answer is that one
can, for sufficiently large mass.  However, one cannot conclude in this case
that the gravitational field must also be quantized, since for masses for
which one can determine the path, the fringe separation in the diffraction
pattern would shrink to below the Planck length, $\ell_{\rm pl} =
(G\hbar/c^3)^{1/2}$, where $G$ is Newton's gravitational constant and $c$ is
the speed of light.  However, position measurements are fundamentally limited
in accuracy to scales $\simge \ell_{\rm pl}$ \cite{calmet}, and thus
distinguishing so a fine pattern cannot be carried out.  Unlike in the
electromagnetic case, where the interference pattern is suppressed due to
decoherence caused by the radiated photons, the pattern in the gravitational
case becomes immeasurably fine, not because the particles radiate quantized
gravitons.

\section{Measurement of the Coulomb field}

    In the experiment sketched in Fig.~\ref{setup} a charged particle enters
the apparatus from the left side, goes through a double slit, and hits the
screen ($b$).  The spacing of the slits is $d$, and $L$ is the distance from
the particle emitter ($a$) to the screen.  The Coulomb field of the electron
is measured at distance $\sim R$ in the plane of the slits, sufficiently far
away from the apparatus that there can be no back-reaction from the distant
measurement of the electromagnetic field.  Thus $R \simge cT$, where $T$ is
the time of the flight of the particle, $\simeq L/v$, with $v$ the particle
velocity.  We consider only non-relativistic particles, in which case the
longitudinal Coulomb field of the electron at distance $R \sim cT$ is larger
than the transverse radiation field by a factor $\sim c/v$.  We assume that
the Coulomb field is determined by the charge in the usual manner.

\begin{figure}[htbp]
\begin{center}
\includegraphics[width=6cm,keepaspectratio=true]{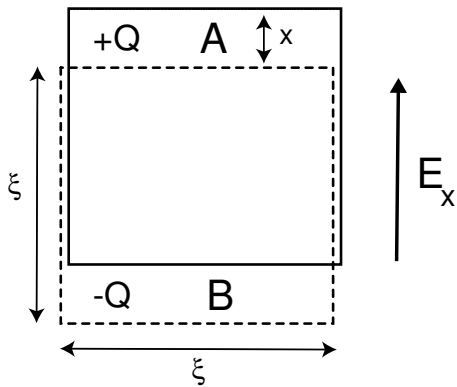}
\end{center}
\caption{Bohr-Rosenfeld apparatus for measuring the electric field.  The
positively charged object $A$ slides on the negatively charged fixed object
$B$.}
\label{apparatus}
\end{figure}

    To distinguish whether the particle goes through the upper or lower slit
one needs to measure the electric field to at least an accuracy $Ze(1/R^2 -
1/(R+d)^2) \sim Zed/R^3$ (with $d\ll R$).  The quantum limit on the
measurability of a weak electric field $E$ was obtained by Bohr and Rosenfeld
\cite{BR1933,BR1950}.  In an early discussion of such a quantum measurement,
Landau and Peierls \cite{LP1931} noted that if one attempts to measure the
field by its effect on a point charge, radiation recoil introduces
uncertainties in the measurement that diverge for short measuring times, and
thus concluded that ``in the quantum range \ldots the field strengths are not
measurable quantities."  To avoid this problem, Bohr and Rosenfeld envisioned
measuring the average of the electric field over a region of space-time, using
an extended apparatus consisting of an object $A$ of mass $M$ and volume $V_A$
with extended charge $Q$, tethered by Coulomb forces to a similar object $B$
with background charge $-Q$.  See Fig.~\ref{apparatus}.  The background charge
is fixed in space, but $A$ is displaced by an electric field from its
equilibrium position.  The apparatus measures the field by detecting the
deflection of $A$ from its equilibrium position.  The net equilibrium charge
density of the apparatus is zero in the absence of an external field that
displaces the object from the background.  In their analysis they first assume
quantization of the electromagnetic field, and show how vacuum fluctuations of
the field in the region limit the accuracy of field measurements.  They then
go on to show that the accuracy of the measurement of a single field is
limited by the uncertainty principle applied to the apparatus, without the
need to invoke field quantization.  We give a schematic derivation of this
result (see also the recent discussions in
Refs.~\cite{compagno,hnizdo,saavedra}.)

    The relative motion of $A$ and $B$ is a harmonic oscillator whose
frequency $\omega$ is readily derived from the familiar expression for the
plasma frequency ($\omega_p^2 = 4\pi n e^2/m$), namely $\omega^2 = 4\pi
Q^2/MV_A$.  When $A$ is displaced relative to $B$ by a distance $x$, the
restoring force acting between them is
\beq
    F = -M\omega^2 x = -4\pi Q^2x/V_A.
\eeq
Thus, an external field $E_x$ acting on $A$ for time $T'$ changes the
momentum of $A$ by $p_x = (E_x Q - 4\pi Q^2 x/V_A)T'$, from which one would
deduce an electric field,
\beq
 E_x = 4\pi Q x/V_A + p_x/QT'.
 \label{force}
\eeq

    Since $p_x$ and $x$ obey the uncertainty relation, $\delta x \delta p_x
\simge \hbar $, we see from minimizing the right side of Eq.~(\ref{force})
with respect to $\delta x$ that the uncertainty in the measurement of $E_x$ is
independent of $Q$, and given by the Bohr-Rosenfeld relation, $\delta E_x
\sim \sqrt{\hbar/V_AT'}$.  For simplicity we assume cubic geometry of $A$
and $B$, with $V_A = \xi^3$, The measurement time $T^\prime$ is at most the
time
of flight, $T$, since further increasing the measurement time does not help to
distinguish the paths; thus we take $T^\prime = T$.  In addition the length
$\xi$ of interest is at most the Coulomb pulse width, $cT$, since
a longer size does not help to distinguish the paths either.  With $\xi = cT$,
we obtain the limit of accuracy of the measurement of the Coulomb field:
\beq
 \delta E_x \sim \sqrt{\frac{\hbar}{\xi^3 T}}\,\, .
\eeq

    To estimate the critical scale of charge of particles above which one
begins to be able to distinguish the path, we take the measuring apparatus to
be located from $R$ to $R + \xi$ above the upper slit.  Then, when a particle
with charge $Ze$ passes through the upper slit, the average Coulomb field in
the apparatus is
\beq
    \frac{1}{\xi}\int_0^{\xi} \frac{Ze}{(R+x)^2}dx = \frac{Ze}{R(R+\xi)}.
\eeq
Similarly, the average electric field when the particle passes through the
lower slit is $Ze/(R+d)(R+d+\xi)$, where $d$ is the slit interval.  Hence
to distinguish the paths the apparatus needs to distinguish an electric field
difference
\beq
  \Delta E =
   \frac{Ze(2R+\xi)}{R^2(R+\xi)^2}d,
\eeq
a decreasing function of $\xi$.  Since to measure the path, one needs
$\Delta E > \delta E$ (the measurement uncertainty), or
\beq
  Ze \simge \frac{R^2(R+\xi)^2}{d(2R+d)} \sqrt{\frac{\hbar}{\xi^3 T}}.
\eeq
With $\xi \sim R \sim cT$ we find that the scale of critical charge $Z_1$
above which one can begin to distinguish the path is
\beq
  Z_1 \simeq \frac{1}{\sqrt{\alpha}}\frac{cT}{d},
\eeq
where $\alpha = e^2 / \hbar c$ is the fine structure constant.  Note that
$Z_1\gg1$, so that one could never detect the path with ordinary electrons or
other particles of charge $\sim |e|$.  For illustration, from the parameters
corresponding to the experiment of Ref.~\cite{tonomura}:  $d \sim 1$ $\mu$m,
and $cT \approx 6$ cm, we estimate $Z_1 \simeq 7\times 10^5$.

    One can in fact, for general $Z$, determine partial information on the
paths, the amount of information increasing with $Z$.  Writing $p(D_u,l)$ as
the probability of the particle having taken the lower path and the detector
detecting it to have taken the upper path, $p(D_u,u)$ as the probability of
the particle having taken the upper path and the detector detecting it to have
taken the upper path, etc., one can quantify the information in terms of the
{\em distinguishability} $\mathcal{D}$
\cite{wooterszurek,jaeger,englert,jacques}
\beq
  \mathcal{D} &= \left| p(D_u,u) - p(D_l,u) \right| + \left| p(D_l,l) -
  p(D_u,l) \right|.
\eeq
Since $p(D_u,u) + p(D_l,u)+ p(D_l,l) + p(D_u,l)=1$, $\mathcal{D}\le
1$.

    To calculate $\mathcal{D}$ we note that the detector determines the
electric field through simultaneous measurement of the position and momentum,
which leads to a Gaussian uncertainty of width $\delta E$ in the measured
value of the electric field from the expected value.  For the particle taking
the upper path, producing an expected (averaged) electric field $E_u$ at the
detector, the probability distribution of the measured electric field is
\beq
  P_u(E) = \frac{1}{\sqrt{2\pi} \delta E} e^{-(E-E_u)^2/2\delta E^2},
\eeq
with a similar expression for the field distribution $P_l(E)$ for the
lower path in terms of the expected $E_l$.  Since $E_u > E_l$, we can for
simplicity regard the detector as having detected the particle taking the
upper path if the measured value of the electric field is greater than $(E_u +
E_l)/2$, and as having taken the lower path otherwise.

    With the assumption that the amplitudes for the particle taking the upper
and the lower paths are equal in magnitude, which is true if the two slits are
located symmetrically,
then
\beq
  p(D_u,u) = \frac{1}{2}\int_{(E_u + E_l)/2}^\infty P_u(E)dE = \frac12 -
 p(D_l,u),
\eeq
with similar equations for $p(D_l,l)$ and $p(D_u,l)$.  With $\Delta E =
E_u - E_l$, the distinguishability becomes
\beq
\mathcal{D} = \frac{1}{\sqrt{\pi}}\int_{-\Delta E / 2\sqrt{2} \delta
E}^{\Delta E / 2\sqrt{2} \delta E} e^{-x^2} dx = \mathrm{erf}(Z/2\sqrt2 Z_1),
\eeq
where $\mathrm{erf}(x)$ is the error function.  We plot $\mathcal{D}$ in
Fig.~3 below for the parameters of Ref.~\cite{tonomura}.

\section{Loss of interference}

    We turn now to the question of how for sufficiently large charge (which
should be $\simle Z_1$) the interference pattern fades out.  The basic
physics is that the particle radiates when being accelerated by the slits, and
undergoes a random change in its phase because it is coupled to a dynamical
degree of freedom, the quantized radiation field.  We do not take into account
any quantum degrees of freedom associated with the slits, i.e., we assume that
they act effectively as a potential on the electron.  The pattern on the
screen is proportional to $\sum_f\left(|\beta_u(b,f)+\beta_l(b,f)|^2\right)$
where $\beta_u(b,f)$ is the amplitude for the particle to go through the upper
slit to point $b$ on the screen, with the electromagnetic field going from its
initial state $|0\rangle$ (the vacuum) to final multi-photon state
$|f\rangle$, and $\beta_l(b,f)$ is the amplitude for the particle to take the
lower trajectory.

    The interference pattern thus has the relative intensity,
\beq
  I(b) = \frac{2{\rm Re}\sum_f\left(\beta_l(b,f)^*\beta_u(b,f)\right)}
 {\sum_f\left(|\beta_u(b,f)|^2+|\beta_l(b,f)|^2\right)}.
\eeq
Although it is possible to carry out a full quantum calculation of the
radiation emitted in diffraction, its essential features are brought out if we
make the simplifying assumption that the charged particle follows a single
straight trajectory along either the upper or lower path from the emission
point $a$ to a given point $b$ on the screen (see Fig.~1). and thus
the emitted radiation has only the effect of changing the phase of the
electron amplitude.  Then
\beq
  \beta_u(b,f) \simeq \langle f|U_u|0\rangle  \beta_u^0,
  \label{beta}
  \eeq
where $\beta_u^0$ is the simple quantum amplitude in the absence of the
electromagnetic field, and
\beq
  U_u= \left(e^{(iZe/\hbar c)\int d{\vec\ell}
  \cdot \vec A(\vec r,t)}\right)_+,
  \label{U}
  \eeq
where $\vec A(\vec r,t)$ is the electromagnetic field operator, and the
integral is time ordered (denoted by the subscript ``+") along the
path.  From Eq.~(\ref{beta}),
\beq
  \sum_f|\beta_u(b,f)|^2 = \langle U_u^\dagger U_u\rangle |\beta_u^0|^2
     =|\beta_u^0|^2,   \nonumber \\
  \sum_f|\beta_l(b,f)|^2 = |\beta_l^0|^2,
\eeq
and
\beq
   \sum_f \beta_l(b,f)^* \beta_u(b,f) = \langle U_l^\dagger U_u\rangle
     {\beta^0_l(b)}^{*} \beta^0_u(b),
 \label{uu}
\eeq
where the brackets denote the electromagnetic vacuum expectation value.
Thus
\beq
  \langle U_l^\dagger U_u\rangle = \big\langle\left(e^{(iZe/\hbar c)\int
   d{\vec\ell} \cdot \vec A(\vec r,t)}\right)_c\big\rangle,
\eeq
where the subscript $c$ denotes the time ordering of the contour integral
from emission to the screen along the upper path and then negatively
time-ordered from the screen back to the emission point along the lower path.
This expression is the expectation value of the Wilson loop around the
path $u-l$ \cite{wilson}.  Since the free quantum electromagnetic field is
Gaussianly distributed in the vacuum,
\beq
 \langle U_l^\dagger U_u\rangle =
   e^{-(Z^2\alpha/2\hbar c)\Phi_{u-l}},
\eeq
where
\beq
  \Phi_{u-l} = \big\langle\left( \oint_{u-l} d\vec\ell \cdot \vec A(\vec r,t)
  \right)_c^2\big\rangle.
  \label{Phi}
\eeq
Writing
\beq
 \langle U_l^\dagger U_u\rangle = {\cal V}e^{-i\zeta},
\eeq
where the {\it visibility} ${\cal V}=|\langle U_l^\dagger U_u\rangle|$ is
$\le 1$, and the phase shift $\zeta$ is real, we have
\beq
   I(b) =  \frac{2{\rm
    Re}\left(\beta^0_l(b)^*\beta^0_u(b)e^{-i\zeta}\right)}
 {\left(|\beta^0_u(b)|^2+|\beta^0_l(b)|^2\right)}{\cal V}.
  \label{vis}
\eeq

    The coupling to the radiation field reduces the intensity of the
interference pattern by ${\cal V}$, as well shifting it via $\zeta$.  By
symmetry, the shift vanishes at the center point on the screen (and is
otherwise not relevant to the present discussion).  Since the Coulomb field
does not enter the states of the radiation field in ${\cal V}$,
Eq.~(\ref{vis}) gives a valid description of the interference pattern whether
or not an attempt is made to distinguish paths by detecting the Coulomb field
at large distances.

    The real part of $\Phi_{u-l}$, entering the visibility, is given by the
same integrals as in Eq.~(\ref{Phi}) without time ordering along the contour,
since $\vec j(\vec r,t)\cdot \vec A(\vec r,t)$ is Hermitian \cite{ford1}:
\beq
 {\rm Re}\,\Phi_{u-l} =
 \big\langle\left(\oint_{u-l} d\vec\ell \cdot \vec A(\vec r,t)
\right)^2\big\rangle.
\eeq
To estimate the visibility we write the free electromagnetic field
operator in terms of photon annihilation and creation operators:  $\vec A(r,t)
\simeq \sum_k \sum_{\lambda_k} (2\pi \hbar c/k\Omega)^{1/2}(a_k \vec\lambda_k
e^{i(\vec k\cdot \vec r - \omega t)} +h.c.)$, where the $\vec \lambda$ are the
photon polarization vectors, $\omega = ck$, and $\Omega$ is the quantization
volume.  For non-relativistic motion ($v\ll c$) along a classical trajectory,
\beq
  {\rm Re}\,\Phi_{u-l}
  =  \int \frac{\hbar c d^3k}{(2\pi)^2 k}\sum_{\lambda_k}\left|\oint_{u-l}
   dt e^{- i\omega t} \vec \lambda_k\cdot \vec v(t)\right|^2 \nonumber \\
\hspace{36pt}   =\frac{2\hbar c}{3\pi}  \int kdk \left|\oint_{u-l}
  dt e^{-i\omega t}\vec   v(t)\right|^2.
 \label{oint}
\eeq
With the simplifying assumption that on the upper path the velocity
undergoes a sudden change at the slits, from $\vec v_1$ to $\vec v_1\,'$ (see
Fig.~1), and from $\vec v_2$ to $\vec v_2\,'$ through the lower
slit, then in the limit of large time of passage, $\omega T \gg 1$,
\beq
  \oint_{u-l} dt e^{- i\omega t}\vec v(t)
  = \frac{i}{\omega} (\vec v_1 -\vec v_1\,' - \vec v_2 + \vec v_2\,'),
  \label{oint1}
\eeq
For $\omega \simle 1/T$, the integral is proportional to $T$.  Near
the center of the pattern, $\vec v_2\,'\simeq \vec v_1 $ and $\vec v_1\,'
\simeq \vec v_2$, so that
\beq
\left|\oint_{u-l} dt e^{- i\omega t}\vec v(t)\right|^2
  \simeq \frac{4}{\omega^2} (\vec v_1 - \vec v_2)^2,
\eeq
and
\beq
\log {\cal V} \simeq -\frac{4 Z^2 \alpha}{3\pi c^2}(\vec{v}_1 - \vec{v}_2)^2
   \int_{1/T}^{\omega_{\rm max}} d \omega \frac{1}{\omega} .
\eeq
The integral over $\omega$, nominally logarithmically divergent at large
$\omega$, is physically cut off by $\omega_{\rm max}$, the maximum frequency
of emitted photons, which from energy conservation cannot exceed $mv^2/2\hbar
= \pi v/\lambda$, where $\lambda$ is the de Broglie wavelength of the
interfering particle.  The lower cutoff is effectively $1/T$; hence
\beq
    \log {\cal V} \simeq -\frac{4 Z^2\alpha}{3\pi c^2}(\vec{v}_1 -
    \vec{v}_2)^2 \log (\pi L/\lambda).
  \label{logv}
\eeq
Equation~(\ref{logv}) is essentially the non-relativistic limit of the
result of Ref.~\cite{breuerpetruccione}.  For $L \gg d$, $(\vec v_1-\vec
v_2)^2 \simeq (2d/T)^2$, and finally, \cite{footnote4}
\beq
    {\cal V}\simeq \exp \left\{ - Z^2 \frac{16\alpha}{3\pi}\frac{d^2}{(cT)^2}
    \log (\pi L/\lambda)  \right\},
\eeq

    Since the path length must be many de Broglie wavelengths, the charge
$Z_2$ above which the visibility becomes less than $1/e^2$ obeys,
\beq
    Z_2 \simeq \frac{cT}{d\sqrt\alpha} \frac{1}{[\log (\pi L/\lambda)]^{1/2}}
    < \frac{cT}{d\sqrt\alpha} \simle Z_1.
\eeq

\begin{figure}[htbp]
\begin{center}
\includegraphics[width=8cm,keepaspectratio=true]{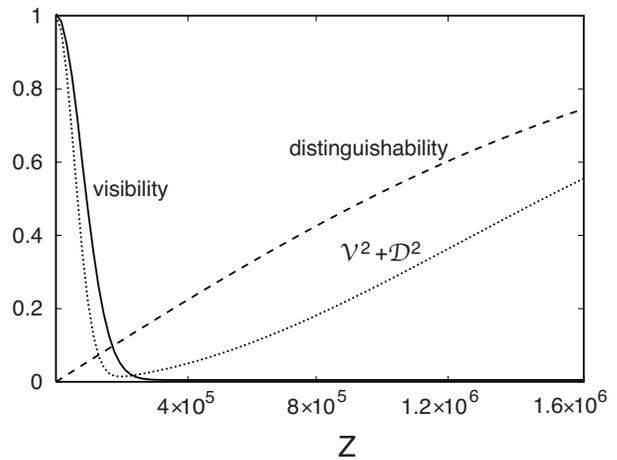}
\end{center}
\caption{Visibility and distingishability vs charge for the parameters of
Ref.~\cite{tonomura},
for which the characteristic charge $Z_1$ for distinguishing paths by
measuring the Coulomb field is $\sim 7\times 10^5$, and the characteristic
charge $Z_2$ for loss of interference is $\sim 1.5\times 10^5$.  Also shown
is $\mathcal{V}^2 + \mathcal{D}^2$ (dotted line).}
\label{vis1}
\end{figure}

    The visibility and distinguishability are closely related; as $Z$
increases the interference pattern fades away on the scale $Z_2$, while the
distinguishability of the paths by measurement of the Coulomb field grows on
the scale $Z_1$.  Quantitatively,
\beq
 \mathcal{V}^2 + \mathcal{D}^2 =
 \exp\left(-\frac{32}{3\pi}\left(\frac{Z}{Z_2}\right)^2\right)
   + \mathrm{erf}\left(\frac{Z}{2\sqrt2 Z_1}\right)^2
 \nonumber \\   \equiv f(Z).
\eeq
Since $f(0) = f(\infty) = 1$, and for $Z_2 < 8Z_1/\sqrt3$ and
$0<Z<\infty$, $f(Z) < 1$, namely
\beq
\mathcal{V}^2 + \mathcal{D}^2 \le 1,
\eeq
in agreement with the inequality derived by Jaeger et al.  \cite{jaeger}
and Englert \cite{englert}.  Figure \ref{vis1} shows the visibility and
distinguishability as functions of $Z$, as well as $\mathcal{V}^2 +
\mathcal{D}^2$, for the parameters of the experiment
of Ref.~\cite{tonomura}, given above.  With these parameters, $\log (\pi
L/\lambda)\sim 20$.

    A simple interpretation of the decrease in visibility, in terms of the
Aharonov-Bohm effect \cite{aharonov-bohm}, is that the closed electron loop,
$u-l$, encircles a fluctuating electromagnetic field which shifts the
interference pattern randomly, thus tending to wash it out.  The
interpretation of the reduction of the pattern in terms of a random flux
requires photon emission processes, and is equivalent to the present
discussion.  Indeed, for the subset of processes in which there is no photon
emission, the modification of the interference pattern is given by $\langle
U_l^\dagger\rangle \langle U_u\rangle$ [cf.~(\ref{uu})], where the brackets
denote states with zero photons.  Now
\beq
  {\rm Re}\log \langle U_u\rangle =
     -\frac{Z^2\alpha}{3\pi}  \int kdk \left|\oint_{u}
  e^{-i\omega t}\vec   v(t)\right|^2 \simeq \frac14 \log {\cal V};
 \label{uint}
\eeq
the reduction reflects the loss of forward-scattering amplitude owing to
photon emission processes.  Thus, the zero-photon emission pattern is
multiplied by a factor ${\cal V}^{1/2}$; the suppression of the zero-photon
pattern at charge $\sqrt2 Z$ equals the suppression of the total visibility at
charge $Z$.  The phase of $\langle U_l^\dagger\rangle \langle U_u\rangle$ is
essentially proportional to the difference of real parts of the electron
self-energy corrections on the upper and lower paths, corrections that do not
contribute to the diminution of the interference pattern.

\section{Measuring the path by gravity}

\begin{figure}[htbp]
\begin{center}
\includegraphics[width=9cm,keepaspectratio=true]{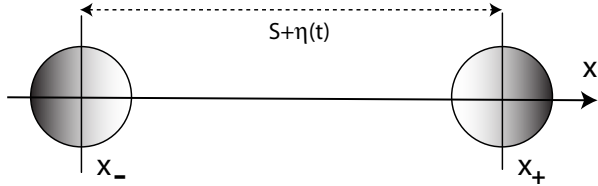}
\end{center}
\caption{Gravitational field detector}
\label{gravitational_detector}
\end{figure}

    Finally, we ask if it is possible to detect the path by measuring the
fluctuations in the (Newtonian) gravitational potential at large distance as a
particle of sufficiently large mass passes through the slits.  In this
scenario, the Newtonian gravitational field plays the role of the Coulomb
field for charged particles.  We consider detecting the change of the
Newtonian gravitational field by using a modern gravity wave detector, e.g., a
highly sensitive laser interferometer \cite{HRS} (a measurement not
equivalent to detecting possible gravitational radiation produced by the mass
going through the slits).  Figure \ref{gravitational_detector} sketches such
a detector.  As before, the x-axis lies in the plane of the slits.  We assume
that the mirrors in the detector are tied down in the lab frame; to a first
approximation, the distance between the mirrors (or equivalently the ends of a
Weber bar) is a harmonic degree of freedom, with oscillator frequency,
$\omega$ (which includes the gravitational attraction of the two mirrors).

    We derive schematically the response of the detector to a Newtonian
gravitational potential $\phi (x,t)$.  In the presence of $\phi$, the
positions of the mirrors, $x_\pm$, obey the Newtonian equations of motion,
\beq
\frac{\partial^2 x_\pm}{\partial t^2} &
  = \mp\frac12 \omega^2 [x_+(t)-x_-(t)-S] -\phi^\prime (x_\pm),
 \label{eom}
\eeq
with $S$ the equilibrium distance between the mirrors, and the prime
denoting differentiation with respect to $x$.  We write $x_\pm = x_0 \pm
(S+\eta)/2$, where $x_0$ is the midpoint between the mirrors in equilibrium,
and $\eta$ is the relative displacement of the mirrors caused by the
gravitational pulse.  Then linearizing in $\eta$ and $\phi''$ we have
\beq
  \frac{\partial^2 \eta(t)}{\partial t^2}
  = - \omega \eta(t)  - \phi^{\prime \prime}(x_0) S.
\eeq
For simplicity we assume that $\phi$ is zero before the gravitational
pulse reaches the detector, and is constant in time during the detection.
With initial conditions $\eta(0) = \eta^\prime (0) = 0$, we obtain
\beq
    \eta(t) = -\phi^{\prime \prime}(x_0)S \frac{1 - \cos \omega t}{\omega^2}.
 \label{etaphi}
\eeq

    The accuracy required for the measurement of $\phi^{\prime \prime}(x_0)$
is
\beq
\Delta\phi^{\prime \prime}(x_0)
    = 2Gm\left(\frac{1}{R^3} - \frac{1}{(R+d)^3}\right) \sim \frac{Gmd}{R^4},
 \label{acc-phi}
\eeq
where $m$ is the mass of the particle, and the measuring apparatus, as
before, is at a distance $R$ from the slits.  Thus, since $1 - \cos \omega t
\le (\omega t)^2/2$, one needs to measure $\eta$ to an accuracy,
\beq
 \Delta \eta \simle \frac{Gmd ST^2}{R^4} < \frac{Gmd}{R c^2},
\eeq
which implies that the mass scale for which one can begin to distinguish the
path obeys,
\beq
  \frac{Gm^2}{\hbar c} \simge \left(\frac{\Delta \eta}{\ell_{\rm pl}}\right)^2
     \left(\frac{R}{d}\right)^2.
\eeq
Physically the uncertainty $\Delta \eta$ must exceed the Planck
length \cite{footnote5}, and thus
\beq
  \frac{Gm^2}{\hbar c} \simge  \left(\frac{R}{d}\right)^2;
  \label{mass}
\eeq
the mass scale must be a factor $R/d$ larger than the Planck mass,
$\sqrt{\hbar c/G }\sim 2 \times 10^{-5}$ g. For $R/d\sim 6\times 10^4$
\cite{tonomura}, the scale would have to be of order 1 g.

    The interference pattern caused by a particle whose mass obeys the
condition (\ref{mass}) has a fringe separation, \beq \delta_f \sim
\frac{L}{d}\frac{\hbar}{mv} \simle \ell_{\rm pl}\frac{cT}{R} \simle \ell_{\rm
pl}, \eeq which implies that when the mass is large enough to allow which-path
detection via gravity, the pattern becomes immeasurably fine, of order the
Planck length or shorter.  This result assures the consistency of quantum
mechanics; however, unlike in the electromagnetic case, consistency does not
require that the gravitational field be quantized \cite{footnote6}. (Although a decrease of the visibility of the
pattern
would arise were gravity quantized, as in the electromagnetic situation,
detailed calculations of the diminution would depend on the detailed theory of
quantized gravity assumed, an issue we do not address here.)

    In summary, when one can distinguish the path of a particle by measuring
the electromagnetic or gravitational field at large distance, interference
disappears.  For large enough charge on the interfering particle, emission of
quantized electromagnetic radiation destroys the interference, while for large
enough mass, the pattern becomes too fine to be discerned.

\begin{acknowledgments}
    This research was supported in part by NSF Grants PHY05-00914 and
PHY07-01611.  We are grateful to Professors A. Aspect, T. Hatsuda, P.~Kwiat,
N.D.  Mermin, B. Mottelson, and C.J.~Pethick for helpful discussions, and to
the Niels Bohr Institute and Aspen Center for Physics where parts of this work
were carried out. 
\end{acknowledgments}


\begin{thebibliography}{99}
\bibitem{footnote4} Note that emission of photons
with wavelengths $\lambda$ larger than the slit width $d$ contributes to the
decrease in visibility, even though such photons give little or no information
about the path.  The reason is that photon emission leads to fragmentation of
the total amplitude, $\sum_n \sum_{f_n} \left(C_{f_n}^u|u,f_n\rangle +
C_{f_n}^l|l,f_n\rangle\right)$, among photon states $f_n$ with various numbers
of photons, $n$.  Here $\sum_n \sum_{f_n}|C_{f_n}^{u}|^2= \sum_n
\sum_{f_n}|C_{f_n}^{l}|^2 =1$.  Only states $|u,f_n\rangle$ and
$|l,f_n\rangle$ with the same photon state can interfere; the total weight of
the interfering terms $|\sum_n \sum_{f_n}C_{f_n}^{l*} C_{f_n}^u|$ must be $\le
1$.
\bibitem{footnote6} As in the
electromagnetic case, one expects a crossover with increasing mass from
indistinguishable to distinguishable paths.  However, a better understanding
of the nature of space-time on the Planck scale is required to determine a
quantitivative visibility.
\end{thebibliography}
\end{document}